\documentclass{aa}  
\usepackage{graphicx}
\usepackage{txfonts}
\begin{document}
\title{A new (2+1)D cluster finding algorithm based on photometric redshifts:
large scale structure in the  Chandra Deep Field South}


\author{Dario Trevese \inst{1}, Marco Castellano \inst{1}, Adriano Fontana\inst{2}\and Emanuele Giallongo\inst{2}}

   \offprints{Dario Trevese \email{dario.trevese@roma1.infn.it}}

   \institute{Dipartimento di Fisica, Universit\'{a} di Roma ``La Sapienza'', P.le A. Moro 2, 00185 Roma \\               
         \and
             INAF - Osservatorio Astronomico di Roma, Via di Frascati 33, 00040 Monte Porzio Catone\\
             }

   \date{Received/Accepted}

  \abstract
   {}
{We study galaxy clustering and explore the dependence of galaxy properties on the  the environment up to a redshift z $\sim$1, 
on the basis of a deep multi-band survey in the Chandra Deep Field South.}
   {We have developed a new method  which combines galaxy angular positions and photometric
  redshifts to estimate  the local galaxy number-density. This allows both the detection of overdensities
in the galaxy distribution and the study of the  properties of the galaxy population as a function of the
environmental density.
}
   {We detect two moderate overdensities at z$\sim$0.7 and z$\sim$1 previously identified spectroscopically. 
We find that the fraction of red galaxies within each structure
increases with volume density, extending to z$\sim$1 previous results. We measure ``red sequence'' slopes 
consistent with the values found in X-ray selected clusters, supporting the notion
that the mass-metallicity relation hold constant up to z$\sim$1.}
   {Our method based on photometric redshifts  allows to extend structure detection
and density estimates up to the limits of photometric surveys, i.e.considerably deeper than spectroscopic surveys.
Since X-ray cluster detection at high redshift is presently
limited to massive relaxed structures, galaxy volume density based on photometric redshift appears as
a valuable tool in the study of galaxy evolution. }

\authorrunning{D. Trevese et al.}
\titlerunning{large scale structure in the CDFS}
\maketitle

\section {INTRODUCTION}
In recent  years deep multi-band  galaxy surveys, together with the  development of
reliable  population  syntheses, and  models  for  galaxy formation/evolution
which incorporate feedback  from the galactic environment,
 have determined a dramatic increase of our understanding of the
observed {\it average}  evolution of the  galaxy population in  cosmic time.
On the other end, clusters and groups of galaxies are the ideal laboratories 
where it is possible to study the environmental effects, which  cause the {\it differences }of
the distribution of galactic types from the {\it average} (Carlberg et al. \cite{carlb}, Gomez et al. \cite{gomez}).
For this reason finding
overdense regions at  high redshift  plays a  central role in 
understanding galaxy formation and evolution.
Unbiased  samples of  clusters  and  groups of galaxies can  be  obtained from  3D
spectroscopic surveys, where the ability of  a Friend  of Friend (FOF) method (Huchra \& Geller \cite{huchra82}) 
to find over-densities is  only limited  by the uncertainty caused by the
galaxy  velocity  dispersion. However, spectroscopic surveys
reach limits of about 5 magnitude brighter than photometric ones, as e.g. in the SDSS survey 
(York et al. \cite{york}), so that complete spectroscopic galaxy redshift surveys of large areas of the sky
are not available at the boundary of the visible Universe.
To minimize the effect of the background/foreground 
objects in the detection of galaxy overdensities,   two-dimensional (2D)  surveys,  
which  are  deeper, require
additional  {\it a  priori}  assumptions on  either galaxy  luminosity
function (LF), as in the Matched Filter algorithm (Postman et al. \cite{postman}) or the presence
of a  red sequence (Gladders et al. \cite{gladd}). Biases produced  by these
assumptions can hardly be evaluated for at high redshift.
Multi-band photometric surveys are able to provide redshifts for large statistical samples of
galaxies at the deepest observational limits, though with a lower accuracy 
($\sigma_z \approx$ 0.05) as compared with spectroscopic ones ($\sigma_z \approx $0.001)
(Fontana et al. \cite{font}, Bell et al. \cite{bell}).
Various algorithms exist to evaluate photometric redshifts. They can rely on the position in a multi-dimensional
colour space respect to galaxies of known redshift (Connolly et. al. \cite{connolly}). Alternatively it is possible to fit the observed
SED with either empirical (Lanzetta et al. \cite{lanz96}) or theoretical templates (Giallongo et al. \cite{giallo98}).
Photometric redshifts have been successfully used to trace the evolution in cosmic time
of the {\it average} galaxy population (Poli et al. \cite{poli}, Giallongo et al. \cite{giallo}). 
However, the present accuracy of photometric redshifts  leaves no room for standard FOF methods.
In fact,  the linking distance $\delta_{FOF}$  must  be larger than the 
distance uncertainty  $\delta_{phot}$ associated with the photometric redshift,
otherwise even close neighbours, which on average appear at a r.m.s. distance of the order of $\delta_{phot}$,
fail to be linked.
On the other hand, $\delta_{FOF}$
must be smaller than the average galaxy-galaxy  distance $d_g$, to avoid a percolation of the ``friendship'' path
through the galaxy distribution leading to huge unphysical chains of galaxies linked by the FOF algorithm,
even in the absence of any group or cluster structure: thus the condition 
$\delta_{phot} < \delta_{FOF} < d_g $ must hold.
However, the minimum uncertainty  attained by  photometric redshifts is $\Delta z  \approx 0.05 \cdot (1+z)$,
corresponding to  $\delta_{phot} \ge $180 Mpc for z $\ge$ 1, while 
the average number density of bright galaxies is $\sim 10^{-2} Mpc^{-3}$, corresponding to an
average galaxy-galaxy distance  as small as $d \approx$ 5 Mpc (Bahcall \cite{neta}), i.e.
we have $\delta_{phot} >> d_g $.
This has been discussed in detail by Botzler et al. \cite{botz} who proposed an extended friends-of-friends method (EXT-FOF)
which applies a FOF method to angular distances, in redshift slices which are defined on the basis of photometric redshifts,
taking into account their intrinsic uncertainties.
In the present paper we present an alternative approach to the use of photometric redshifts.
Our  algorithm evaluates the 3D galaxy density
using angular positions and photometric redshifts with the purpose of: 
i)  detecting galaxy over-densities in three dimensions, 
and ii) assigning to each galaxy a measure of the environmental density, to
extend the analysis of the environmental effects
on galaxy evolution to the limits of photometric surveys.
The method is applied to a deep photometric survey of the Chandra Deep Field South (Cimatti et al. \cite{cima02}),
where we re-discover some previously known overdensities and we find a clear relation between 
the fraction of blue and red galactic types and the local density.
The paper is organised as follows. In section 2 we present a new method to define a 
galaxy- volume density using angular positions and photometric redshifts.
In section 3 we apply the algorithm to a deep photometric survey of a portion of
the Chandra Deep Field South to identify some high-$z$ clusters and analyse the
dependence of their galaxy population on local density. 
In section 4 we study the evolution 
in cosmic time of the red sequence.
In section 5 we check the reliability of the results
and possible extrapolation to deeper surveys by applying
the algorithm to a simulated galaxy catalogue. 
In section 6 we discuss the results and section 7 provides a summary. 
Throughout the paper we adopt consensus cosmological parameters $\Omega_{\Lambda}$=0.7, $\Omega_M$=0.3, $H_0$=70 km s$^{-1}$ Mpc$^{-1}$.

\section {(2+1)-DIMENSIONAL DENSITY FROM PHOTOMETRIC REDSHIFTS}

As far as background/foreground effects are acceptable, namely at
relatively low redshifts, the surface density is sufficient to detect galaxy clusters
or to study their morphology,  as in the case of Abell (\cite{abell}) or Zwicky (Zwicky et al. \cite{zwi}) clusters 
or in the later works of the Edinburgh-Durham Cluster Catalogue (Lumsden et al. \cite{lum}).
A surface density  
\begin{equation}
 \Sigma_n = n/(\pi D_{p,n}^2) 
\end{equation}
 of galaxies can be 
computed considering the projected distances
$D_{p,n}$ to the n-th nearest neighbour, and can be used to study 
the effect of the environment on the distribution of galactic types, as 
done by Dressler (Dressler \cite{dress1}, Dressler et al. \cite{dress2}).
Similarly, when spectroscopic redshifts are available, 
is possible to compute the three-dimensional (non projected) distance $D_n$ to the n-th nearest neighbour and derive a volume density 
\begin{equation}
\rho_n = 3n /(4 \pi D_n^3).
\end{equation}
The method we propose for deep photometric surveys, where spectroscopic galaxy redshifts are not available,
consists in combining, in the most effective way, the   
angular position with the (much less accurate) distance as computed from the photometric
redshift. 
In principle it is possible to compute for each galaxy the distance
from its neighbours from the angular positions and photometric redshifts, once a cosmological model is assumed.
In  practise we prefer to divide the survey volume in cells whose extension in  different directions 
($\Delta\alpha, \Delta \delta, \Delta z$)
depends on the relevant positional accuracy and thus are elongated in the radial direction. 
Then for each point in space (i.e. for each cell) we count neighbouring objects 
of increasing distance, until a number $n$ of objects is reached.
We define the density associated to the cell as $$\rho = n/ V_n$$ where $V_n$ is the volume which
includes the n nearest neighbours. 
The choice of $n$ is a trade off between spatial resolution and signal-to-noise ratio.
In the following the number $n$ is chosen in such a 
way that  $n$ objects are present in a single cell in the regions of maximum density. 
Notice that, in principle, the way of searching neighbouring objects of increasing distance in not univocal,
since distance steps (cell sizes) in different directions can be chosen arbitrarily.
The result will be a different smoothing, and resolution, in different directions. The choice we adopt
is to keep the maximum resolution in transversal and radial directions allowed by the data. 
Since the uncertainty in the
radial direction is much larger, this will correspond to elongated cells and lower radial resolution.
Steps $\Delta z$ in the radial direction smaller than the photometric redshift uncertainty would 
uselessly increase the computing time.
In counting galaxies
we must take into account the increase of limiting luminosity with increasing redshift for a given flux limit.
If $m_{lim}$ is the  limiting (apparent) magnitude in a fixed observing band, 
at each redshift $z$
we detect  only objects brighter than an absolute magnitude $M_{lim}(z)$,
decreasing (brightening) with $z$ . 
We can assume  a reference redshift $z_c$  below which we detect all objects brighter
than the relevant $M_c \equiv M_{lim}(z_c)$, below which we neglect the incompleteness.
At $z > z_c$ the fraction of detected objects is: 
\begin{equation}
s(z)=\frac{\int^{M_{lim(z)}}_{-\infty}\Phi(M)dM}{\int^{M_c}_{-\infty}\Phi(M)dM} 
\end{equation}
where $\Phi(M)$ is the galaxy luminosity function.
Thus, in evaluating the galaxies number density, we apply a {\it limiting magnitude correction}
by assigning a weight $w(z)=1/s(z)$ to each detected galaxy of redshift $z$.
In this way we correct the systematic underestimate of density, caused to the increasing
fraction of galaxies which fall below the brightness limit for increasing redshift. Of course 
the noise in the derived density
increases as the square root of $w(z)$, but the advantage is to obtain 
a density scale independent of redshift, at least to a first approximation.
We stress that, in general, the distance modulus $m-M$  depends not only on the luminosity 
distance $D_L(z)$, and 
thus on the adopted cosmological model, but also on the k- and evolutionary- corrections 
which, in turn, depend on the galactic type. 
Moreover the luminosity function itself depends on both the wavelength $\lambda$ and cosmic time.
However, as far as $s(z)$ is not much smaller than one, it is possible to adopt simple
representations of these effects, and any correction to the limiting magnitude correction will
represent a second order effect.
Practically the choice of the cell sizes is determined by the accuracy of the photometric redshifts.
In fact,
once the cells have transversal sizes which are much smaller than the 
radial one  (say 1/1000), a further increase of the transversal resolution does not justify the corresponding 
increase of the computing time. At least, this is true as far as we do not expect structures which are 
physically strongly elongated in the radial direction.
In the application presented in the following sections, the relative redshift uncertainty is
$\Delta  = |z_{spec}-z_{phot}|/(1+z)\approx 0.05$ (Cimatti et al. \cite{cima02a}),
corresponding to a length scale $l_z \approx 200$,  Mpc at $z \sim 1$. Cells sizes are arbitrary and are chosen
small enough not to degrade the spatial resolution, while avoiding a useless increase of the computing time.
We have adopted $\sim 60 $ Mpc in the radial direction and $\sim 60$ kpc at $z \sim 1$ in transversal direction,
 corresponding to $\Delta \alpha = \Delta \delta = 3.68$ arcsec.
Galaxy clusters or, more generally, galaxy overdensities are defined as connected 3-dimensional 
regions with density exceeding
a fixed threshold. Once overdensities are identified, it is possible to analyse the properties 
of galaxies as a function of the local density.

\section {GALAXY CLUSTERS IN THE CHANDRA DEEP FIELD SOUTH}

The dataset used to identify distant clusters and to study the properties of their member 
galaxies is the deep
photometric catalogue of the K20 survey (Cimatti et al. \cite{cima02}, \cite{cima02a}) containing photometry in the $UBVRIZJK$ bands
of a 6.38x6.13 arcmin field in Chandra Deep Field South (CDFS) (Giacconi et al.  \cite{giacco}).
The sample is limited to $I_{AB}<25$, while the photometric depth in the other bands allows to 
assume that virtually
all galaxies in the catalogue have 8-band photometry, except a few objects with very extreme 
colours. We have added to the spectroscopic redshifts of the K20 all public spectroscopic redshifts
in our field from GOODS-MUSIC catalogue (Grazian et al. \cite{graz} and refs. therein).  
The catalogue contains 1749 galaxies among which 292 have spectroscopic redshifts and the remaining have only photometric redshifts. 
The procedure for deriving photometric redshifts and test their accuracy are described in 
Cimatti et al. \cite{cima02a},where it is shown that the distribution of the  fractional error 
$\Delta  \equiv (z_{spec}-z_{phot})/(1+z_{spec})$  is not gaussian. After excluding ``outlayers'' with $\Delta > 0.15$,  
which represent less than 9\% of the total, we obtain $\sigma_{\Delta}=0.05$.

Figure 1 shows photometric redshifts versus spectroscopic redshifts,
with the  uncertainty $0.05 (1+z)$  indicated by the dashed lines.

\begin{figure}
   \centering
\resizebox{\hsize}{!}{\includegraphics{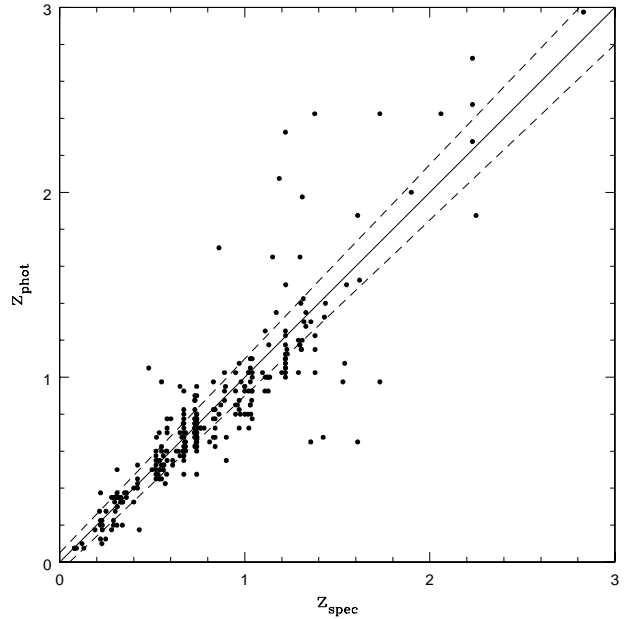}}

      \caption{
Photometric redshifts z$_{phot}$ versus spectroscopic redshifts z$_{spec}$ for all the galaxies in the catalogue
with spectroscopic observations (Grazian et al. \cite{graz} and refs. therein). 
Dashed lines indicate the r.m.s. uncertainty 0.05(1+z).}
         \label{Fig1}
   \end{figure}
We have constructed (2+1)D maps of the volume density $\rho_{n}$, with $n=10$,
computing $s(z)$ in equation (3) on the basis of a cosmologically evolving luminosity function.
This has been taken from Poli et al. \cite{poli}, who derived the rest-frame B luminosity
function which is directly sampled until the rest-frame blue is observed in the K band, namely up to
a redshift of about 3.5. In their analysis Poli et al. \cite{poli} find little density evolution at the faint end with respect to the 
local values, while at the bright end  a brightening increasing with redshift is apparent with respect to the local LF.\\
The choice $n=10$ corresponds to the maximum number of objects in a single cell at high density
as discussed in Section 2.
\begin{figure}
   \centering
\resizebox{\hsize}{!}{\includegraphics{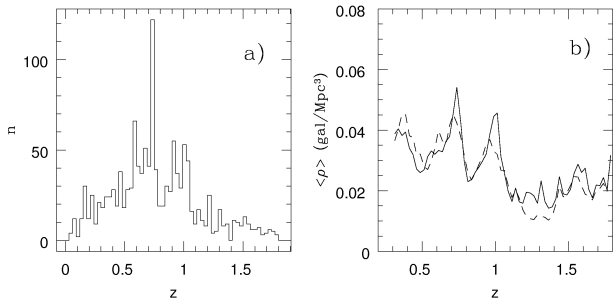}}

      \caption{
a) The distribution of photometric redshifts of the sample.
b) Average $\rho_{10}$ density on the entire field, in redshift bins,  versus z as determined by the (2D+1) algorithm using
for all sources:
i) photometric redshifts (dashed line); ii) photometric redshifts, or spectroscopic redshifts whenever available
(continuous line).
              }
         \label{Fig2}
   \end{figure}
\begin{figure}
   \centering
\resizebox{\hsize}{!}{\includegraphics{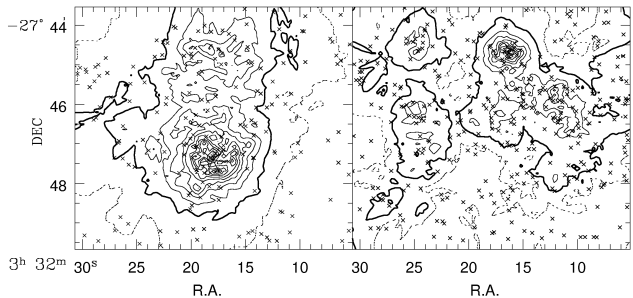}}
      \caption{Isodensity contours of the surface density $\Sigma_{10}$ as 
computed in the redshift slice $0.70 < z < 0.75$, which includes the first detected structure (left panel).
b) Same plot for the second structure in the redshift slice $0.90 < z <1.10$  (right panel).
              }
         \label{Fig3}
   \end{figure}
\begin{table}
      \caption[]{Detected structures.}
         \label{KapSou}
     $$ 
         \begin{tabular}{c c c c}
            \hline
            \noalign{\smallskip}
            \# & $\alpha_{2000}$ & $\delta_{2000}$ & $z$   \\
            \noalign{\smallskip}
            \hline
            \noalign{\smallskip}
1&03 32 20.08& -27 47 07.20& 0.70\\
2&03 32 16.74& -27 45 52.99& 1.00\\
3&03 32 16.51& -27 46 45.95& 1.55\\
            \noalign{\smallskip}
            \hline
         \end{tabular}
     $$ 

   \end{table}

To see how the resolution of photometric redshifts compares with real objects distribution, we show in
Figure \ref{Fig2}  the histogram of photometric redshifts in the field. Two main clumps
appear about redshifts 0.70, 1.00.
A comparison with the distribution of spectroscopic redshifts of the CDFS (see Gilli et al. \cite{gilli} fig. 1) 
clearly indicates the reality of the two clumps,
although the two peaks at z=0.67 and z=0.73 found by Gilli et al. \cite{gilli} in the distribution of spectroscopic redshifts 
are barely resolved in our  photometric redshift distribution.
On the basis of spectroscopic redshifts, Gilli et al.  \cite{gilli} found that both structures are spread over the field,
although the latter includes a cD galaxy, suggesting a dynamically relaxed status (see Sarazin \cite{sar}).

To check to what extent our results depend on the presence of
spectroscopic redshift in our catalogue, we show in Figure \ref{Fig2}b
the average density in the field as a function of redshift, as
computed from photometric redshifts only, or including also
spectroscopic redshifts whenever the latter are available.  The two
curves look similar. In fact using $z_{spec}$ instead of $z_{phot}$
does not change significantly the density, once the cell size has been
chosen on the basis of the (much lower) $z_{phot}$ accuracy. This
means that our results do not rely on the availability of several
spectroscopic redshifts in the field.  Notice that the density
reported in Figure \ref{Fig2}b, which is averaged over the entire
field, is not used to detect structures.  For this purpose we select
individual cells with density above a given threshold, and then we
look for connected volumes.  The choice of the threshold is an
arbitrary trade off between completeness and reliability. From our
numerical simulations (see section 5) we found that a threshold
corresponding to about three to five times the average density can
identify richness zero Abell clusters. More complex simulations would
be necessary to reliably evaluate the degree of contamination as a
function of the richness class and redshift.  In real data, from the
distribution of the densities of individual cells we found that
thresholds of 0.078 Mpc$^{-3}$ or 0.13 Mpc$^{-3}$ (i.e. 3 or 5 times
the average) isolate 2.2\% or 0.5\% respectively of the total cell
number.  To avoid excessive contamination from random density
fluctuations we adopted a threshold $\rho_{10}^{thresh}$ =0.08
Mpc$^{-3}$ which selects about 2.0\% of the cells.  In this way we
identified  three  over-densities listed in table 1.  We
adopted a redshift independent threshold to detect structures of
comparable density at any redshift. In principle, this implies an
higher probability of contamination at higher redshifts: we plan to
explore this issue in future work extending to higher redhifts.

The first structure, at z=0.70 is approximately centred on the cD galaxy,
whose position, in turn, corresponds to the centre of the extended
X-ray source CDFS566 (Giacconi et al. \cite{giacco}).
As noted above, we cannot resolve,  in redshift space, the wall-like structure at z=0.67 described by
Gilli et al. \cite{gilli} which then contaminates the structure at z=0.70. 
In spite of that we find a relatively concentrated structure with full width at half maximum of about 0.12 Mpc.
After the statistical subtraction of the background/foreground field galaxies
the number of galaxies within the Abell radius $R_A$ is $N_c$=182, of which 38
are between $m_3$ and $m_3+2$, corresponding to a richness class 0.
From the spectroscopic redshifts  we can evaluate a
 velocity dispersion along the line of sight, $\sigma_{p}=$ 334 $\pm$ 31 Km s$^{-1}$, 
for the 39 galaxies belonging to the peak in the redshift distribution centred at at z=0.73,
the uncertainty being computed by a bootstrap method. 
The relevant virial mass $M = \frac{3\pi}{2}\frac{\sigma^{2}_{P} R_{PV}}{G}$ is $M=1.19\cdot10^{14}M_{\odot}$,
where  $R_{PV}=N(N-1)/\sum_{i>j}{R^{-1}_{ij}}$ is the projected virial radius 
and $R_{ij}$  are the projected distances between each pair of the N=39 galaxies 
(Heisler et al. \cite{heisler}, Girardi et al. \cite{nico}).
For the sole purpose of displaying the morphology of the density field we show, in Figure \ref{Fig3}a,  the isolines of the 
surface density $\Sigma_{10}$ (see section 2), evaluated  in the redshift slice
0.70$ < z_{phot} <$ 0.75. A similar plot for the overdensity at z$\simeq$1.00 is shown
in Figure \ref{Fig3}b where the galaxy 
number within the Abell radius, with $m_3< m < m_3+2$,  barely reaches the formal threshold of 30 
corresponding to richness class 0, depending on the exact location of the adopted centre.
The spectroscopic redshift distribution suggests the  existence of two distinct peaks at 0.97 and 1.04: the former 
associated with galaxies around the main overdensity and the latter corresponding to the south east extension.
The analysis of possible substructures of this overdensity requires further spectroscopic data.
The third clump we find at $z =1.55$  does not appear in the distribution of spectroscopic
galaxy redshifts which is limited to brighter fluxes respect to our photometric data.
On the other hand,the accuracy of our photometric redshifts is statistically checked against spectroscopic ones only for
$z_{spec} \la 1$, so that further data would be needed to assess the reality of this structure.
However a peak in the distribution of the X-ray selected AGNs in the field is present at about z=1.55.
Thus, as far as we can assume that  distribution of AGNs traces the large scale distribution of matter we can say that we are detecting
a structure  not previously seen in galaxy spectroscopic surveys.
The Abell richness of the third structure at z$\sim$1.55 cannot be evaluated, since $m_3+2$ falls below the limiting magnitude $m_I$=25.
The peak in the photometric redshift distribution  
contains 57 objects spread along a  moderate over-density crossing the field from north-west to south-east
likely related to the above discussed large-scale structure traced by  X-ray selected AGN (Gilli et al. \cite{gilli}).
The association of an environmental density with individual galaxies allows both
a further assessment of the nature of the detected overdensities and the analysis of the relation
between galaxy spectral energy distribution and the environment.
\begin{figure}
   \centering
\resizebox{\hsize}{!}{\includegraphics{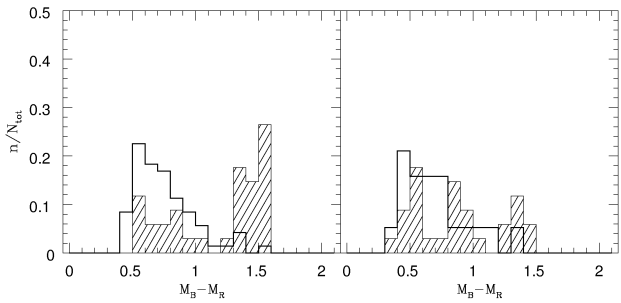}}
      \caption{
Galaxy colour distributions: at high density ($\rho_{10}>0.08 Mpc^{-3}$) (shaded histogram), at low density ($\rho_{10}<0.03 Mpc^{-3}$), for the two structures at z$\sim$0.70 (left) and z$\sim$1.0 (right).
              }
         \label{Fig4}
   \end{figure}

\begin{figure}
   \centering
\resizebox{\hsize}{!}{\includegraphics{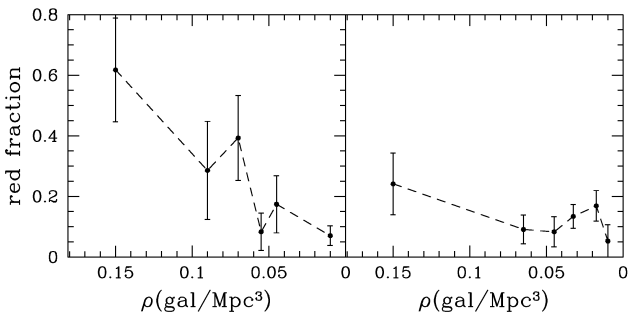}}
      \caption{
Fraction of galaxies with rest frame colour B-R$>$1.25 as a function of 
the volume density $\rho_{10}$, for the two structures at z$\sim$0.70(left) and z$\sim$1.0 (right).
Error bars represent Poissonian fluctuation.
              }
         \label{Fig5}
   \end{figure}

A strong colour bi-modality of the colour distribution has recently been confirmed on the basis of a large
galaxy sample of about 150,000 objects from the Sloan Digital Sky Survey (Strateva et al. \cite{strat} and refs. therein).
This bi-modality has been shown to maintain up to z$\approx$2-3 (Giallongo et al. \cite{giallo}),
with a local minimum in the colour distribution which evolves in redshift and  represents the natural separation between the ``blue'' and 
``red'' galaxy populations, the latter defining an {\it average} red sequence of the field.
Instead, Figure \ref{Fig4} shows rest-frame B-R colour distribution in overdense ($\rho_{10} >$ 0.08 Mpc$^{-3}$) and
underdense ($\rho_{10}<$0.03 Mpc$^{-3}$) regions both at  z$\sim$0.7 and z$\sim$1.
Following Carlberg et al. \cite{carlb}, we conservatively adopt the constant rest frame colour B-R=1.25 as a boundary between the two populations.
The excess of red galaxies in overdense regions respect to the field, clearly appears
in Figure \ref{Fig4}.
According to a Kolmogorov-Smirnov test, the probability of the null hypothesis that the colours inside and outside the overdensities
are randomly drawn from the  same distribution is 4.2$\times$10$^{-7}$ for the cluster at z=0.7 and 0.058
for the cluster at z=1.0.
Due to the insufficient statistics,
a similar colour segregation cannot be detected in the case of the z$\sim$1.55 overdensity,
which will be studied as soon as deeper photometric data will become available (Trevese et al. \cite{tre06}).
In the two former overdensities we can study the fraction of red galaxies as a function
of the density $\rho_{10}$.
The result is shown in Figure \ref{Fig5}, where both clusters show  a  decrease of red fraction as a function of 
$\rho_{10}$, marginally significant at z=1.0 and more evident at z=0.7.
This colour segregation is clearly related to the morphological segregation first found
by Dressler \cite{dress1} for local clusters successively extended to z=0.5  by Dressler et al. \cite{dress2}.
Our analysis of colour segregation allows to extend a quantitative investigation of environmental
effects up to  redshifts where morphological studies become unfeasible.

\section {COSMOLOGICAL EVOLUTION OF THE RED SEQUENCE}

\begin{figure}
   \centering
\resizebox{\hsize}{!}{\includegraphics{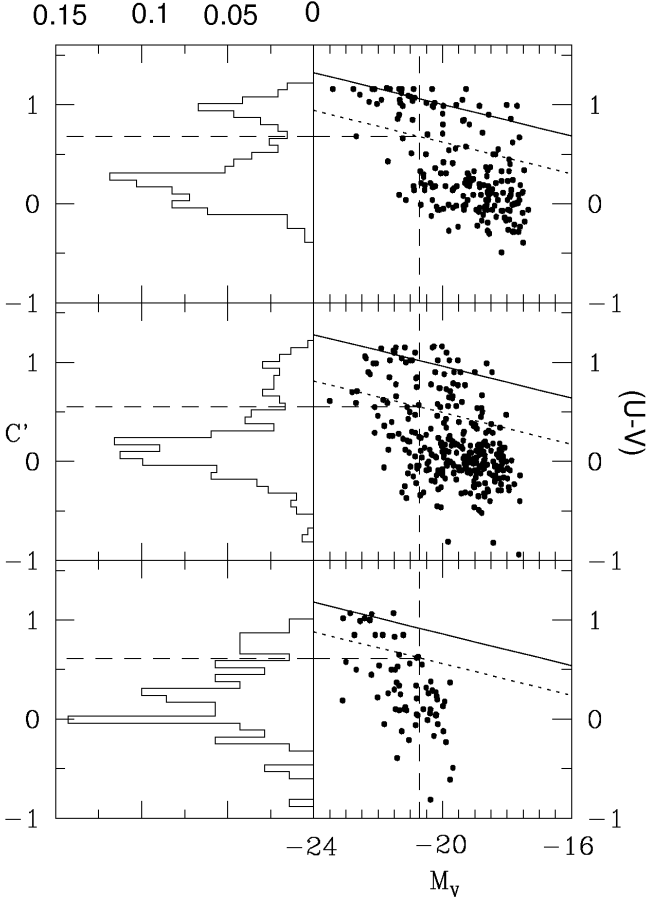}}
      \caption{
Left panels: histograms of the C' colour, defined in the text, for all the objects in the intervals 
$0.7<z<0.8$, $0.9<z<1.1$, $1.4<z<1.7$ (from the top). Dashed lines, corresponding to a local minimum,
 define the red and the blue populations.
Right panels: rest-frame (U-V) vs. $M_V$ diagrams: the continuous line represents the fit
to the points belonging to the red population,  with a fixed slope $\alpha_{RS}$=0.08.
The dashed vertical line represents the reference absolute magnitude $M_V=-20.7$. The dotted line 
indicates the valley separating the two populations.
}
\label{Fig6}

\end{figure}

\begin{figure}
   \centering
\resizebox{\hsize}{!}{\includegraphics{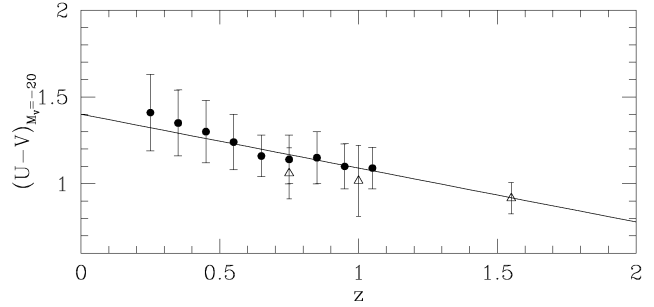}}
      \caption{
The colour  $C'_{RS}$ at $M_{V_{rest}}$=-20 of the {\it average} red sequence, as computed from COMBO17 survey  (circles),
and resulting from the present analysis (triangles). 
The straight line represents a linear fit on all the data analysed in the COMBO17 survey
 (adapted from Bell et al. \cite{bell}).
              }
         \label{Fig7}
   \end{figure}

\begin{figure}
   \centering
\resizebox{\hsize}{!}{\includegraphics{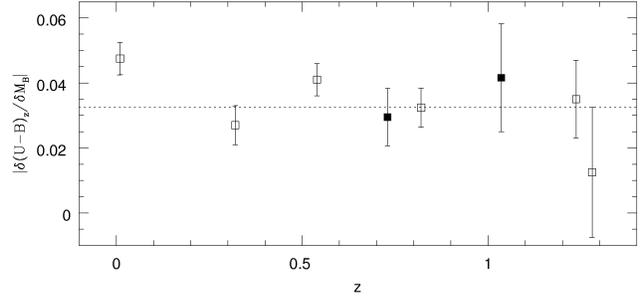}}
      \caption{
The slope of the rest-frame (U-B) vs. M$_B$ in the  galaxy clusters 
collected by Blakeslee et al. \cite{blake} (open squares)
and in the two structures detected in the CDFS (filled squares). 
The dotted line represents the average value derived by Blakeslee et al. \cite{blake}
}
         \label{Fig8}
   \end{figure}

We first study the cosmological evolution of the {\it average} red sequence of the field. 
Following Bell et al. \cite{bell}
it is possible to define for each galaxy a colour index C' reduced to  $M_{V_{rest}}$= -20
by shifting  each galaxy in the C-M diagram to $M_{V_{rest}}$=-20 along the red sequence:
$$C'=C+\alpha_{RS}\cdot (M_{V{rest}}-20)$$
where C is the rest-frame colour, (U-V)$_{rest}$ in our case, and
$\alpha_{RS} \equiv \partial (U-V)_{rest} / \partial M_{V_{rest}}$ is the slope of the red sequence,
In practise, for a direct comparison with Bell et al. \cite{bell}
who adopt a different cosmology, we assume as reference magnitude $M_{V_{rest}}=-20.7$ instead of $-20$.
A distribution of the C', instead of C, colours allows a better identification
of the ``Early type'', or red galaxies, population which defines the {\it average} red sequence. 
Following Bell et al. \cite{bell}
we assume a constant slope 
$\partial (U-V)_{rest} / \partial M_{V_{rest}} = -0.08$
which is 
derived from a sample of nearby clusters (Bower et al. \cite{bower}). This assumption is justified by the analysis
of Blakeslee et al. \cite{blake} who find a constant slope for different galaxy clusters up to z=1.2 (see the next paragraph).
At higher redshifts Giallongo et al. \cite{giallo} find a decrease of the red sequence slope 
$\partial (U-V)_{rest} / \partial M_{B_{rest}}$ from -0.098 to -0.062, in the two wide redshift bins
0.4-1 and 1.3-3.5 respectively. 
In the present analysis, which is limited to z$<$1.7, we neglect this change of slope and 
we estimate the red sequence colours in three relatively narrow redshift bins.
Figure \ref{Fig6} (a-c) shows the
 rest-frame colour-magnitude diagrams for all galaxies in the field, selected 
in the three slices of photometric redshift, $0.7<z<0.8$, $0.9<z<1.1$, $1.4<z<1.7$ 
and the relevant C' distributions.
From the C' colour distribution we select the members of the red population as those galaxies lying redwards of the natural
minimum which separates the blue and red populations. 
A fit in the C-M diagram of the red population with a straight line of fixed slope $\alpha_{RS}$
 defines the colour $C'_{RS}$ of the {\it average} red sequence
in different bins of redshift.
Figure \ref{Fig7}, adapted from Bell et al. \cite{bell} shows the colour $C'_{RS}$  of the  {\it average} 
red sequence in our field in the above redshift intervals , 
compared with the colour of the {\it average} red sequence as a function of redshift obtained from
COMBO17 data, after a correction on $\Delta C'=C'_{combo}-C'_{CDFS}=-0.08$ 
as derived from a comparison of the colours of those high-z galaxies which are in common in the 2 catalogues COMBO17 and K20.
Our error bars represent the  r.m.s. dispersion of C' in each redshift bin. 
We can conclude that
the  $\Delta C'$ vs. z relation in our analysis is consistent with Bell et al. \cite{bell} up to z$\simeq$1, moreover the point at z=1.5
lies on the linear extrapolation of their points.
As a further step, we 
measured the slope of the red sequence in the two clumps at redshifts 0.7 and 1.0.
This has not be done for the third clump at z=1.55, due to insufficient statistics.
We selected galaxies with an environmental density above a threshold $\rho_{10}$=0.08 $gal/Mpc^{3}$ (see section 3).
We then isolated the red population from the histograms of the C' colour which are clearly bimodal.
Finally, we evaluated the slope of the colour-magnitude relation for this population in the
U-B vs. B diagram, as done by Blakeslee et al. \cite{blake}. 
Our results show  that the slope of the red sequence is consistent with being the same in local clusters
and in the two main overdensities (z$\sim$0.7 and z$\sim$1.0).
Blakeslee et al. \cite{blake} found little or no evidence of evolution of this slope,
out to z=1.2. Our results (Fig. \ref{Fig8}) lie within 1$\sigma$ from their average value 
$| < \partial (U-B)_{rest} /  \partial M_{B_{rest}} > | = 0.032$.
According to the standard interpretation (Arimoto \& Yoshii \cite{ari}, Kauffmann \& Charlot \cite{kauf}) this implies that
the mass-metallicity relation holds the same from z=0 up to at least z=1

\section{A CHECK OF THE ALGORITHM ON SIMULATED CLUSTERS} 

\begin{figure}
   \centering
\resizebox{\hsize}{!}{\includegraphics{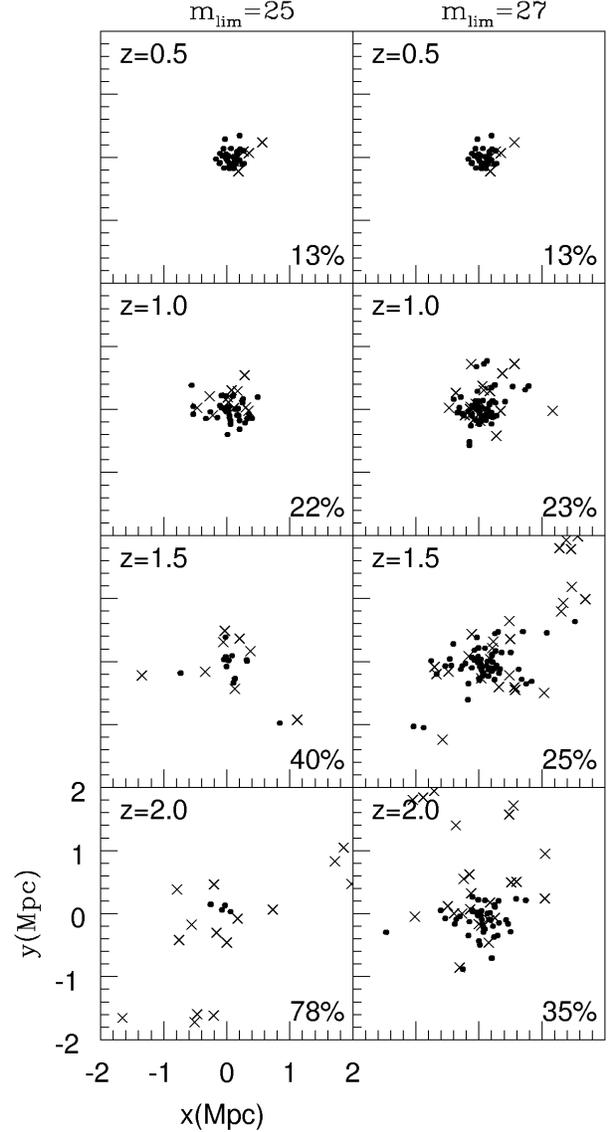}}
      \caption{
Galaxies above an environmental density $\rho_{10}$ threshold 5 times the average, for redshifts
0.5, 1.0, 1.5 and 2.0 and for catalogues with two different limiting magnitudes.
Filled dots represent real members of the simulated richness class 0 clusters, while crosses represent interlopers.
The contamination is reported in each panel.
              }
         \label{Fig9}
   \end{figure}

To check the reliability of the algorithm in detecting clusters of various types and redshifts
we created a series of mock  catalogues including field galaxies and clusters.
Field objects were uniformly distributed on a square of 6 x 6 arcmin centred on the cluster,
with a  space density  and absolute magnitude distributions assigned according 
to the redshift-dependent Schechter-like (Schechter \cite{schecht}) LF derived by Giallongo et al. \cite{giallo}:
\[ \Phi(M,z)dM=0.4ln10\phi^{\ast}(z)[10^{0.4(M^{\ast}(z)-M)}]^{1+\alpha}exp[-10^{0.4(M^{\ast}(z)-M)}], \]
with $\phi^{\ast}=\phi^{\ast}_{0}(1+z)^{\gamma}$ and $M^{\ast}=M^{\ast}_{0}-\delta log(1+z)$.
Clusters are represented by a number-density distribution
 $$n(r)=n_0[1+(r/r_c)^2]^{-3/2}$$  (see Sarazin \cite{sar}) 
with a typical core radius $r_{c} = 0.25  Mpc$.  To take into account the uncertainty on photometric
redshifts,  we assigned to each galaxy a random redshift with a mean corresponding to the cluster
redshift and a dispersion $\sigma_z=0.05$ corresponding to a velocity dispersion along the line of sight
$\sigma_v=15000/(1+z)$ Km s$^{-1}$ (Hogg \cite{hogg}). 
Once  position and redshift are assigned to a galaxy,
an absolute blue magnitude $M_B$ is extracted from
 a Schechter LF, with  parameters
$M_B^{\ast}=-20.04$ and $\alpha=-1.05$ for ellipticals and $M_B^{\ast}=-19.48$ 
and $\alpha=-1.23$ for spirals,
derived by De Propris et al. \cite{depro} from the analysis of 2dF clusters.
The fraction of different galaxy types was chosen according to Goto et al.  \cite{goto}.
Different values of the central density $n_0$
are chosen to produce clusters of various richness classes.
Table 2 reports the values of the parameters for elliptical and spiral galaxies in the field.
\begin{table}
      \caption[]{LF parameters.}
         \label{KapSou}
     $$ 
         \begin{tabular}{c c c c c c}
            \hline
            \noalign{\smallskip}
  type& $\phi^{\ast}_{0}$ & $\gamma$ & $M^{\ast}_{0}$ & $\delta$ & $\alpha$ \\
            \noalign{\smallskip}
            \hline
            \noalign{\smallskip}
Ellipticals&0.0106&-2.23&-20.03&2.72&-0.46\\
Spirals& 0.0042&-0.52&-20.11&2.35&-1.38\\
            \noalign{\smallskip}
            \hline
         \end{tabular}
     $$ 

   \end{table}
To evaluate the total number of cluster and field objects $ N=\int^{M_{LIM}(z)}_{-\infty}\Phi(M,z)dM $ 
we computed the limiting rest-frame magnitude
in the B band given
by:
\[M_{LIM}=m_{l}+C-K-E-25.0-5.0\cdot log_{10}D_L\]
$m_{l}=25$ is the limiting AB  magnitude at the effective wavelength of the  I band, K and E are, respectively, the type
dependent K-correction and evolutionary correction (Poggianti \cite{pogg}), C is the relevant rest-frame (B-I) colour and  
 $D_L$ is the luminosity distance. 
We assumed  a fixed proportion of Sa (70\%) and Sc (30\%)in the field catalogue.
Various mock catalogues were generated at each cluster redshifts 0.5, 1.0, 1.5, 2.0  
corresponding to Abell (\cite{abell}) richness classes 0,1,2,3.
Figure  \ref{Fig9} shows the galaxies belonging to a richness class 0 cluster, above a $\rho_{10}$ density threshold five times the average,
as seen at different redshifts and limiting magnitudes. Real cluster members and interlopers are represented by 
filled dots and  crosses respectively. For a limiting magnitude $m_{lim}$=25, the cluster is detected with an acceptable contamination
up to z=1, while z=1.5 may be assumed as a detection limit. However, in the case of a deeper survey, with   $m_{lim}$=27
comparable with the new GOODS-MUSIC catalogue (Grazian et al. \cite{graz}), whose analysis is in progress, the same 0 richness
cluster is well detected up to z=1.5 and still visible at z=2.  

Clearly the ability of the algorithm in separating the overdensities 
increases with their richness and their
angular distance. 
Thus, to perform a conservative evaluation of the algorithm, we simulated 
two relatively low density structures  perfectly aligned along the line of sight.
We have considered pairs of overdensities,  representing two richness 0 clusters as those described above.
The first is at redshift
z=1.0 and the second at various higher redshifts. 
We found that, depending on the chosen threshold,  the two structures can be separated if the distance  
is greater then $\Delta$z=0.15. However, once N galaxies are assigned to a single
structure, the redshift of the latter can be determined with an accuracy  of about $\Delta z / N^{1/2}$,
which can be an order of magnitude smaller than $\Delta z$.
The above results refer to the possibility of resolving (aligned) structures. 
It is worth noting that in real cases,once a (2+1)D density map is available,
it is possible to adopt a multi-threshold technique, 
to identify possible physical substructures and/or projection effects.

\section {DISCUSSION}
It is clear that many methods exist for detecting clusters, depending on the type of the available data:
from single-band images, where only brightness can be used to complement angular positions in the identification of overdensities
like in matched filter methods (Postman et al. \cite{postman}),
to images in a few bands where colours allow to identify the cluster red sequence (Gladders et al. \cite{gladd}), to 
images in several bands allowing the determination of photometric redshifts, to spectroscopic redshift.
Each method has his advantages and limitations. One of the main issues is our ability to compare
the results of numerical simulations with galaxy catalogues  obtained from different data sets. 
A general discussion of these problems has recently presented by Gal \cite{gal06}.
To establish whether one  method is more or less efficient respect to the another in detecting structures 
is not a straightforward task. 
Given a mock catalogue, each method should be optimised, by a proper choice of the relevant parameters, to obtain
a meaningful comparison. Moreover, mock catalogues obtained by different assumptions on cosmic evolution of the spectral and
clustering properties of galaxies would lead to different optimisations. 
The problem of using the photometric redshifts to identify structures in the galaxy distribution 
has been tackled recently by Botzler et al. \cite{botz} who discuss why standard FOF methods,suited for spectroscopic redshift
surveys, cannot be simply applied to photometric redshift surveys without a proper account of the large inherent uncertainties.
They propose an extended EXT-FOF method which applies a two-dimensional FOF algorithm to slices of the galaxy catalogue, which
are defined by photometric redshifts,taking into account their uncertainty. 
Clearly photometric redshifts add crucial information for the identification of structures in the galaxy distribution
through the fitting of the observed spectral energy distribution to either theoretical or observed galaxy templates, 
evolving in cosmic time.
Here we propose and discuss a different way of using  photometric information: instead of comparing the distance between galaxy pairs,
our method uses the statistical information about {\it how many} galaxy are in the neighbourhood of a given point and estimates a local
density without introducing a fixed smoothing scale.
A comparison of our method with EXT-FOF or other methods is  beyond the aim of the present work and
will be the subject of future investigations.
What we want to stress here are only the new informations we obtain with our method:
i) the spatial resolution is maximised in each point, since near density peaks the volume taken 
into account to evaluate the density is small, while it becomes large where the density is low;
ii) once the density map is available, it can be analysed in different ways without re-running the
algorithm. For instance the map can be sliced at different density thresholds to see how structures
which appear separated at high density merge at lower densities. 
In principle it is also possible to apply an harmonic analysis to density maps.\\

We have also obtained some new results
by  applying our (2+1)D method to real data in the Chandra Deep Field South which is one of the most studied fields in the sky.
The existence of deep X-ray data from the Chandra observatory provided a sample of Active Galactic Nuclei (AGN) in the field (Giacconi et al. \cite{giacco}).
The spectroscopic follow up has shown the existence of large-scale structure (Gilli et al. \cite{gilli}).
AGN, thanks to their intrinsic luminosity and 
to strong emission lines, are the ideal tracers of the mass distribution at the maximum distance reachable by optical spectroscopy,
under the assumption that their spatial distribution mimics that of normal galaxies. 
The same depth is either unreachable or requires huge exposure times for normal galaxies. 
Photometric redshifts,though with the limitation imposed by their poor accuracy, permit to identify overdensities of normal galaxies at 
the depth of AGN samples. This allows to study the relation between the space distribution of AGNs and galaxies.
In the case of the CDFS, our photo-z based density detects 
two clusters, at z$\sim$0.7 and z$\sim$1, 
already identified in the spectroscopic redshift survey of Gilli et al. \cite{gilli}. A third peak at z$\sim$1.5 in the 
galaxy density distribution likely corresponds to a peak which is found in the spectroscopic redshift distribution of AGNs but not
of galaxies. The analysis of deeper photometric data, which is in progress (Trevese et al.\cite{tre06}), 
is necessary to confirm the reality of this structure.
Once  galaxies belonging to a cluster are identified,  it is possible to 
study the galaxy type distribution as a function of the local density. 
This provides an evidence of environmental effects on galaxy evolution and a way to quantify these effects. 
In the case of our analysis this allowed to 
prove, in a cluster of redshift 0.7, that the fraction of red galaxies increases with density, as already known
for lower redshift structures (Dressler \cite{dress1} and Dressler et al. \cite{dress2}). 
A similar effect appears at redshift 1, though the evidence in this case is marginal.
Clearly the density computed by our method is not the ``real local density'', 
due to the strong smoothing in the z direction caused by the redshift uncertainty and to the assumptions adopted to 
compensate the increasing loss of faint galaxies at high redshift. Thus any cosmological application
of our method, devoted to study the evolution in cosmic time of galaxy clusters properties and environmental effects, 
will require a calibration through a detailed comparison  with simulated catalogues. 
Obviously the same is true for any other cluster finding technique.
However, there are properties of the cluster galaxy population which, to a first approximation, do not need a comparison with
mock catalogues to provide significant physical information. This is the case of the slope of the cluster red sequence
which has been traced up to z=1.27 (Blakeslee et al.\cite{blake} and refs. therein).
According to Kodama et al. \cite{koda98} the evolution of the colour-magnitude (C-M) slope 
depends on relative age variations in early-type
galaxies with different luminosities. Accordingly, the constancy of the slope up to redshift 1.27 is consistent
with passive evolution of an old stellar population that was formed at high redshift. In this scenario, changes of the C-M relation slope
are expected at redshifts approaching the star formation phase. 
An alternative interpretation of the constancy of the C-M relation,
in the framework of hierarchical models of galaxy formation, has been proposed by Kauffmann \& Charlot \cite{kauf}.
In this scenario metals, once formed,  are more easily ejected from smaller disks. Large (bright) ellipticals 
are more metal-rich because they are formed from the mergers of large disks. 
In selecting rich clusters at high redshift one is biasing samples towards objects that merged at the highest
redshifts and for this reason they {\it appear} to follow the passive evolution. 
In both scenarios the C-M relation is due to a mass-metallicity relation and the
evolution of the C-M slopes at high redshift contains critical informations
on the origin of the C-M slope itself. 
Our analysis supports the evidence of a constant C-M slope up to z=1, meaning a correspondingly constant mass-metallicity
relation.
We stress that most of the highest redshift clusters detected so far were selected in the X-ray band,
in particular those at z=1.24  (Blakeslee et al. \cite{blake}, Rosati et al. \cite{rosati}) and z=1.4 (Mullis et al. \cite{mullis}).
However, both this clusters have a velocity dispersion of about 800 km s$^{-1}$ and intracluster gas 
temperature kT $\sim$ 6 Kev,typical of rich clusters (Bahcall \cite{neta}, Arnaud et al. \cite{arnaud}).
The X-ray luminosity of these two clusters  in the  0.5-2 Kev band are $1.9 \times 10^{44}$  and  $3.0 \times 10^{44}$
erg $ s^{-1}$ $ h_{70}^{-2}$  respectively (Rosati et al. \cite{rosati}, Mullis et al. \cite{mullis}), 
again typical of clusters with richness class greater than 2 (Ledlow et al. \cite{ledlow}).
Rich clusters of these redshifts are just at the limit of our present (2+1)D analysis based on photometric observations.
Work is in progress (Trevese et al. \cite{tre06}) to extend the study to a deeper samples ($K_{AB}$ $\sim$ 27)
which will allow  the analysis at  z$\sim$1.5 of
even poor structures of the type of those detected by photometric redshifts in the present work in the CDFS. 
These less pronounced structures are hardly detectable in X-rays, due to the strong 
dependence on richness of the X-ray luminosity (Ledlow et al. \cite{ledlow}), and their C-M slope could differ
from that of the richest clusters, if it constancy is only apparent and mainly produced by the cluster selection bias
towards richer structures.
  
\section{CONCLUSIONS}

We have presented a new method to  detect local overdensities in the galaxy distribution,
based on: i)  photometric redshifts, with proper account of the
relevant uncertainty, to evaluate distance; ii) distance to the n-th neighbour to evaluate densities.
From a methodological point of view, we can conclude that:
\begin{itemize}
\item{
although  our (2+1)D method is limited by the redshift uncertainty, so that in principle
 only structures separated by more than $\delta_{phot} $ can be resolved, 
it still allows to dramatically increase our possibility of {\it detecting} structures respect to 2D analyses,as shown 
by the example in Figure \ref{Fig9}}
\item{
 if we restrict to structures of the type of Abell clusters, whose 
number density is about $10^{-5} Mpc^{-3}$ (Bahcall \cite{neta}),
the average inter-cluster distance results $d_{cl}\sim$100 Mpc , namely  $\delta_{phot}  \le d_{cl} $,
meaning that  the use of photometric redshifts allows to {\it resolve} even aligned structures, as  shown by 
the simulation in section 5.}
\end{itemize}

From an astrophysical point of view, our analysis of deep multiband photometry in the CDFS field allows the following conclusions:
\begin{itemize}
\item{
we have detected two structures at redshifts 0.7 and 1.0, whose existence was known from previous spectroscopic studies;}
\item{ 
a third structure at redshift 1.5 has also been detected but requires deeper data for confirmation;}
\item{
 the fraction of red galaxies inside the structure at z=1 indicates a  marginal density dependence
while in the structure at z=0.7 the increase of the red fraction with density is seen very clearly;
this extends the results of Dressler et al. \cite{dress2}, Carlberg et al. \cite{carlb} and Tanaka et al. \cite{tana};}
\item{
our results add new evidence in favour of constant slope of the C-M relation in clusters, at least up to z=1
 implying a constant mass-metallicity relation, according to the standard interpretations;}
\item{
our analysis shows that the average C-M relation 
for galaxies belonging to the red population is consistent with a linear  extrapolation of the
relation found by Bell et al. \cite{bell} up to z=1.5, complementing the results 
of Giallongo et al. \cite{giallo}, who have shown that the 
blueing of  the average colour of the red galaxies extends to z$\approx 2-3$;} 
\item{the use of photometric redshifts will allow to analyse the redshift dependence
of the C-M relation at high redshift, even in moderate overdensities,  providing constraints on the very origin of the C-M relation.}
\end{itemize}
Thus, in spite of the low resolution in distance, (2+1)D analysis based on photometric redshift
is an extremely valuable tool to complement other cluster finding techniques and perform large scale structure
studies based on photometric surveys which, at present, posses unique capabilities 
in combining depth and field width.

\begin{acknowledgements}
We thank the anonymous referee for valuable comments and suggestions.
We are grateful to Andrea Grazian and Sara Salimbeni for providing advise and assistance in the use of
photometric and spectroscopic catalogues.
\end{acknowledgements}

\end{document}